\documentclass[12pt,preprint]{aastex}
\begin{document}
\def\kms    {\ifmmode{{\rm ~km~s}^{-1}}\else{~km~s$^{-1}$}\fi}
\def\lo     {~$L_{\odot}$}
\def\mo     {~$M_{\odot}$}
\def\etal   {{ et~al.}}
\def\rmaa   {Rev. Mexicana Astron. Astrofis}
\title{VLA and BIMA observations toward the exciting source of the massive
HH~80-81 outflow}

\author{Y. G\'omez, L.F. Rodr\'\i guez}
\affil{Centro de Radioastronom\'\i a y Astrof\'\i sica, UNAM, Apdo. Postal 3-72 (Xangari) 
              58089 Morelia, Michoac\'an, M\'exico\\
{\tt y.gomez@astrosmo.unam.mx}; {\tt l.rodriguez@astrosmo.unam.mx}}

\author{J. M. Girart}
\affil{Departament d'Astronomia i Meteorologia, Universitat de Barcelona, 
Av. Diagonal 647, 08028 Barcelona, Catalunya, Spain\\
{\tt jgirart@ub.edu}}

\author{G. Garay}
\affil{Departamento de Astronom\'\i a, Universidad de Chile, Casilla 36-D, Santiago, 
Chile\\
{\tt guido@das.uchile.cl}}

\and

\author{J. Mart\'\i}
\affil{Departamento de F\'\i sica, Escuela Polit\'ecnica Superior, 
Universidad de Ja\'en, Calle Virgen de la Cabeza, 2, E-23071 Ja\'en, Spain\\
{\tt jmarti@ujaen.es}}

\begin{abstract}

We present high angular resolution VLA and BIMA observations of NH$_3$, HCO$^+$, 
HCN and SO molecular emission and 1.4, 3.5 and 7 mm continuum emission toward the 
exciting source of the HH~80-81 system. This object is one of the few massive 
protostars known to be driving a collimated outflow. We report the first detection 
of SO 5$_5$--4$_4$ molecular emission toward the exciting source of HH 80-81, 
suggesting that this transition may be a good tracer of molecular gas near massive 
protostars. We also detected toward this source dust continuum emission at 1.4 and 
3.5 mm. From the SO molecular emission and the dust emission we roughly estimated 
that the molecular mass associated with the circumstellar surroundings of the 
exciting source of the thermal jet is in 
the range 1 -- 3 M$_\odot$. Weak and broad (2,2) ammonia emission was also found in 
the direction of the jet suggesting the presence of small amounts of molecular gas 
at high temperatures ($>$ 50~K).

The VLA observations show the presence of three ammonia components toward the HH 
80-81 region. The brightest component peaks at $\sim$8$^{\prime\prime}$ to the NE of 
the thermal jet and is associated with the H$_2$O maser spots in the region. A 
second ammonia clump is located about 25$^{\prime\prime}$ to the NE of the jet and 
is associated with class I methanol masers. The third ammonia component is located 
1$^\prime$ to the south from the thermal jet and may be a molecular core yet without 
stellar formation. The BIMA observations show that the strongest emission in the 
HCO$^+$ and HCN lines originate close to the H$_2$O maser, and cover the same spatial 
region and velocity range as the brightest ammonia component.

\end{abstract}  


\section{Introduction}

The HH~80-81 system,  also known as GGD~27 (Gyulbudaghian et al. 1978), consists of a 
highly collimated radio jet that powers an outflow with a total projected extent 
of 5.3 pc, at an adopted distance of 1.7 kpc (Rodr\'\i guez~et al. 1980; Mart\'\i , 
Rodr\'\i guez, \& Reipurth 1993; 1995; 1998). The thermal radio jet associated with 
HH~80-81 terminates to the south in the extremely bright optical HH objects 80 and 81 
(Reipurth \& Graham 1988; Mart\'\i, Rodr\'\i guez, \& Reipurth 1993, 1995; Heathcote, 
Reipurth, \& Raga 1998) and to the north in the radio source HH~80 North (Mart\'\i 
~\etal~1993; Girart et al.~1994; 2001). The central radio continuum source has a 
bright far-infrared counterpart, IRAS~18162$-$2048, which implies the presence  of 
a luminous ($\sim$ 2 $\times 10^4$~L$_{\odot}$) young star or cluster of stars (Aspin 
\& Geballe 1992). Water maser emission
has been reported in the vicinity of the central radio continuum source
(Rodr\'\i guez \etal ~1978; 1980). The H$_2$O masers do not coincide with the 
jet core, suggesting that  they are being powered by a different star 
(G\'omez, Rodr\'\i guez, \& Mart\'\i ~1995; Mart\'\i, Rodr\'\i guez, \& 
Torrelles 1999). Methanol emission at 95 GHz has been reported by 
Val'tts et al. (2000) to the northeast of the exciting source of the jet at
the position of $\alpha$(2000)=  18$^h$ 19$^m$ 12${\rlap.}^s$89; 
$\delta$(2000)= $-$20$^\circ$ 47$^\prime$ 13${\rlap.}^{\prime\prime}$6 with
a positional error of 10$^{\prime\prime}$. These methanol masers are classified
as class~I which are believed to trace high-mass protostars that have not
significantly ionized their environment.

The most remarkable characteristic of the HH~80-81 system is the high degree of
collimation of the jet, with an opening angle of $\sim$1$^\circ$ (Mart\'\i , 
Rodr\'\i guez, \& Reipurth 1993).
Collimated outflows and accretion disks are thought to be present at the 
same evolutionary stage in the formation of both low-mass and high-mass stars
(e.g. Garay \& Lizano 1999; Kurtz 2000). An accretion disk around the central star 
is believed to be responsible for the ejection and collimation of the gas.  
In contrast with low-mass stars, where disk structures have been observed commonly 
around young objects (Beckwith \& Sargent 1996; Wilner, Ho, \& Rodr\'\i guez 1996; 
Rodr\'\i guez, et al. 1998; Wilner et al. 2000), in high-mass stars it has been 
difficult to establish the presence of accretion disks. Whether the massive 
protostars are formed by accretion (as it is the case for low mass stars) or by 
merging of protostellar condensations (Bonnell \& Bate 2002) remains controversial.  
It has been argued, for example, that the ionizing flux present in massive stars 
could rapidly disrupt the disk. However, the presence of well collimated outflows in 
a few objects also points to the presence of accretion disks surrounding massive stars. 
If massive OB stars are formed by accretion, we expect that disks and jets will be 
present in their earliest stages of evolution, as in the case of low mass stars.
Accretion disks and collimated outflows around B-type protostars have been tentatively 
reported toward Cepheus A HW2 (Rodr\'\i guez et al. 1994; Torrelles et al. 1996), 
IRAS 20126+4104 (Cesaroni et al. 1997; Zhang, Hunter \& Sridharan 1998), G192.16-3.82 
(Shepherd et al. 1998; Devine et al. 1999; Shepherd, Claussen, \& Kurtz 2001), 
AFGL~5142 (Zhang et al. 2002) and AFGL~2591 (Trinidad et al. 2003).  
A possible collimated outflow associated with a 
luminous O-type protostar is IRAS~16547$-$4247, a bright infrared source that  
is associated with a triple radio source consisting of a compact central object and 
two outer lobes (Garay et al. 2003). There is, however, no known unquestionable case 
of a circumstellar disk associated with a massive O-type protostar.  

The luminosity of HH~80-81 is $\sim$2$\times$10$^4$ L$_\odot$, equivalent to that of a
B0 ZAMS star.
Since HH~80-81 possesses an extraordinary well collimated jet, it seems reasonable to
propose that it should be originating from an accretion disk. However, until now
there has been no clear evidence for the existence of this disk. A bipolar CO outflow has
been detected and mapped in the direction of HH~80-81 by  Yamashita ~et al.~(1989) which 
spatially coincides with the infrared reflection nebula in the region. Dense molecular gas
traced by NH$_3$ (1,1) emission was reported by Torrelles et al. (1986) from single dish 
observations, whereas Girart et al. (1994) reported the detection of (1,1) and (2,2)
inversion transitions of ammonia downstream in HH~80 North.
Also a CS molecular clump has been observed towards the exciting source of HH~80-81
which exhibits an elongated structure ($\sim$60 $\times$ 25$^{\prime\prime}$)  
perpendicular to the CO molecular outflow, that has been interpreted as a molecular disk 
(Yamashita ~et al. 1989). However, high angular resolution observations are needed in 
order to establish the existence of a more compact circumstellar disk. We consider 
HH 80-81 as one of the best candidates for the detection of a protoplanetary disk 
using molecular lines. Accordingly, the main goal of this work was to make observations, 
in several molecular lines and with adequate angular resolution, towards 
the core of HH~80-81 to search for an accretion disk  that could account for the 
collimated outflow.

\section{Observations}

\subsection{VLA}

The 1.3 cm line and continuum observations were made on 1999 April 12 with the 
Very Large Array (VLA) of the NRAO\footnote{The National Radio Astronomy
Observatory is operated by Associated Universities, Inc., under
cooperative agreement with the National Science Foundation}\ . 
The array was in the D configuration giving an angular resolution  of 
4$^{\prime\prime}$ at 1.3~cm. Structures larger than 60$^{\prime\prime}$
were undetectable. We observed the (J,K)=(1,1) and (2,2) inversion 
transitions of the 
ammonia molecule in the spectral line mode towards the core of the
HH~80-81 complex. We assumed rest frequencies of 23694.496 and 23722.633 
MHz for the (1,1) and (2,2) transitions, respectively. We used a 
bandpass of 3.125 MHz, centered at an LSR velocity of 10 km~s$^{-1}$ 
and 63 spectral channels 48.8 kHz wide each ($\sim$ 0.6 ~km~s$^{-1}$ at 
the observing frequencies) plus a continuum channel containing the central  
75$\%$ of the total band. 
Flux densities were bootstrapped from observations of 1331+305, which 
was assumed to have a flux density of 2.4 Jy at 1.3 cm. The bandpass calibrator 
was 1229+020 with a flux density of 28.6 Jy and the phase calibrator was 
1733-130 with a flux density of 3.8 Jy at 1.3 cm. 
The data were edited and calibrated in the standard manner using the 
Astronomical Image Processing System (AIPS) developed by the NRAO. 

All spectra were Hanning smoothed to a velocity resolution of 1.2 km~s$^{-1}$.
The $rms$ noise levels in a single spectral channel, after Hanning smoothing,
were 2.5 and 2.2 mJy beam$^{-1}$ (corresponding to 0.38 and 0.33~K) for the (1,1) 
and (2,2) observations, respectively. The system temperature during observations was in the
order of 150~K. 
We applied a primary beam correction  (with the task PBCOR of AIPS) to the
ammonia data in order to have a good estimate of the flux for the emission
that is displaced from the phase center.

\subsection{BIMA}

The observations were carried out in 1999 September 27 at 3.5~mm and 2001 May 6
at 1.4~mm with the 10-antenna BIMA array\footnote{The BIMA array is operated by
the Berkeley-Illinois-Maryland Association with support from the National
Science Foundation.} in the C configuration.  The phase calibrator used was
1733-130.  Absolute flux calibration was performed by observing Mars. System
temperatures were in the 200--400~K range at 3.5~mm and in the 1600-4000~K
range at 1.4~mm.  The 3.5~mm frequency setup included the HCO$^+$ 1--0 and HCN
1--0 lines, whereas the 1.4~mm setup included the SO 5$_5$--4$_4$ and H$_2$CO
3$_{0,3}$--2$_{0,2}$ lines.  Self-calibration was performed in the 1.4~mm data
by using the continuum emission. Maps were made with the  $(u,v)$ data weighted
by the associated system temperatures.  More details of the observations are
given in Table 1.

\section{RESULTS}

\subsection{VLA and BIMA continuum emission}

Using the VLA we detected an unresolved ($\leq 2{\rlap.}^{\prime\prime}6$) 1.3 cm 
continuum source coincident in position with the exciting source of the thermal jet. 
The estimated flux density  at 1.3~cm is 6 $\pm$ 1~mJy. This is consistent with the value 
reported by Mart\'\i , Rodr\'\i guez \& Torrelles (1999).
The 7 mm flux density was obtained from data taken with the VLA on
1994 April 15. The unresolved ($\leq 0{\rlap.}^{\prime\prime}4$) 7~mm source is shown in Figure 1.
The continuum images at 3.5 and 1.4~mm made with BIMA are shown on Figure~2. 
To discuss the nature of this emission we present in Figure 3 the continuum spectrum 
towards the exciting source of the thermal jet at cm and mm wavelengths.
The 20, 6, and 2 cm flux densities are taken from Mart\'\i\ et al. (1993), while
the 1.3 cm data point is that presented above from our VLA observations.
The flux densities  at 7, 3.5 and 1.4~mm toward the exciting source of
the outflow are listed in Table 2.  From Figure 3 we can see that the cm emission shows
the characteristic flat spectrum of nearly optically thin
free-free emission (S$_\nu \sim \nu^{0.2 \pm 0.1}$), but that at 7 mm and shorter wavelengths, 
the flux density rises rapidly. This spectrum suggests that the flux densities at 3.5 and 1.4~mm
are completely dominated by the dust (at 3.5~mm the expected contribution from the
thermal jet is only $\sim$0.01 Jy). The spectral index derived from the 7, 3.5 and 1.4~mm fluxes is
2.6, consistent with optically-thin dust emission. There is also 1.4~mm emission associated
with the C ammonia component (see Figure~2, and Table~3).

\subsection{VLA spectral line emission}

The (1,1) and (2,2) main inversion transitions of ammonia were strongly detected in 
emission toward the northeast of the jet, but only very weak (2,2) 
emission was observed at the position of the exciting source of the thermal jet.

Figure~4 shows the velocity-integrated ammonia (1,1) emission toward HH~80-81.
We identify three main ammonia structures; the bright, C component which exhibits
an extended structure and peaks close to the position of the water masers in the region,
a small ammonia clump to the northeast of the
C component (hereinafter called the NE component)  
and the S component which is located $\sim$1$^\prime$ to the south of the C
component.

Figure~5 shows the channel maps of the NH$_3$~(1,1) and (2,2) emissions
toward the surroundings of the exciting source of
HH~80-81. In general, the ammonia emission from
the molecular cloud in both transitions appears distributed to the northeast
of the thermal jet, spanning over the same velocity range, from
$\sim$ 10 to 14 km~s$^{-1}$. As noted before,
the peak emission of the C ammonia component is nearly 
coincident with the  position of the H$_2$O maser spots (G\'omez, Mart\'\i , \& 
Rodr\'\i guez 1995; Mart\'\i , Rodr\'\i guez \& Torrelles 1999).

Figure~6 presents a close-up of the velocity-integrated emission in the (1,1) and 
(2,2) ammonia lines from the C component and the (1,1) line from the S 
component. The top panel also shows the (1,1) ammonia emission from the NE ammonia 
clump where methanol maser emission has been reported (Val'tts et al. 2000). 
The S component was detected only in the (1,1) ammonia transition and in the 
velocity range from 11.9 to 13.7 km~s$^{-1}$, suggesting it corresponds 
to a cold molecular clump.  

The spectra of the NH$_3$ (1,1) and (2,2) inversion transitions from each individual 
ammonia component (integrated over their whole area of emission) 
are presented in Figure~7. The (1,1) main and satellite lines were detected toward 
each individual ammonia component, while in the (2,2) ammonia transition emission
was detected only in the main line and toward the C and NE components. 
In Table~3 we list the ammonia parameters of the main and satellite  ammonia lines
from the C, NE and S components determined by fitting a Gaussian profile to the
corresponding spectrum.
The peak intensity of these three ammonia
components are around the LSR velocity of +12 km~s$^{-1}$.

\subsection{BIMA spectral line emission}

Figure 8 shows line profiles and maps of the SO 5$_5$--4$_4$, HCO$^+$ 1--0 and HCN 1--0
emission. The SO emission arises from a compact region around the thermal jet. 
A Gaussian fit to the SO profile gives a peak flux density of 7.6 $\pm$ 
1.6~Jy, a central velocity of V$_{LSR}$= 12.1 $\pm$ 0.2 km~s$^{-1}$ and a linewidth of
$\Delta$V= 1.3 $\pm$ 0.4 km~s$^{-1}$.
This is the first time that SO emission is detected toward the exciting source of 
HH 80-81 and suggests that this transition may be a good tracer of molecular gas 
near massive protostars. The HCO$^+$ and HCN emissions trace basically the same 
spatial region as the ammonia central component and in a similar velocity range. 
Weak HCO$^+$ emission is observed toward the NE ammonia clump, confirming its existence.
There is some enhancement of the HCO$^+$ and HCN emission  toward the direction of the
thermal radio jet, but it is difficult to separate the molecular gas arising
from the jet region from the more extended central component. 
Despite the similitude of the HCO$^+$ and HCN integrated maps with the 
integrated ammonia maps, the observed HCO$^+$ and HCN spectra show a clear self-absorption 
feature at the center of their lines (see Fig.~8), which is not observed in the
ammonia lines. It is possibly a true self-absorption since the shortest baselines
also show this absorption feature.  

We also searched for emission in the H$_2$CO 3$_{0,3}$--2$_{0,2}$ line toward the 
region. The integrated velocity spectrum in a velocity range from 10 to 14 km~s$^{-1}$ 
did not show emission at a 3-$\sigma$ level of $\sim$2 Jy~beam$^{-1}$.

\section{DISCUSSION}

The line data presented here reveal the presence of three main molecular structures 
that we called the C, the NE, and the S components. In what follows we discuss 
each of these components and derive their bulk physical parameters, and conclude with 
a discussion of the molecular and dust emission toward the surroundings of the jet.

\subsection{The C component}

This large structure of molecular gas has a peak position located about 
8$^{\prime\prime}$ northeast from the exciting source of the thermal jet. Emission 
toward this component was detected in the (1,1) and (2,2) transitions of NH$_3$, 
and HCN (1--0) and HCO$^+$ (1--0) lines, indicating the presence of considerable 
amounts of cold molecular gas. The strongest H$_2$O maser in the region (marked 
with a triangle in Figures 4 to 8; with a V$_{LSR}$ = $-$61 km~s$^{-1}$ and a peak 
position at $\alpha$(2000)= 18$^h$ 19$^m$ 12${\rlap.}^{s}$509; 
$\delta$(2000)= $-$20$^\circ$ 47$^\prime$ 27${\rlap.} ^{\prime\prime}$42, G\'omez et 
al. 1995), lies close (2$^{\prime\prime}$) to the molecular peak emission of the 
C component (see Table~3). The LSR velocities of the individual H$_2$O maser 
spots are between $-$50 to $-$72 km~s$^{-1}$ (Mart\'\i ~et al. 1999), considerably 
blueshifted with respect to the NH$_3$, HCN and HCO$^+$ molecular emission 
(+12 km~s$^{-1}$). A deep 3.5~cm radio continuum map reveals the presence of a weak 
($\sim$0.05 mJy) unresolved ($\leq$ 0${\rlap.}^{\prime\prime}$2) radio source (VLA~3 
in G\'omez et al. 1995; $\alpha$(2000)=  18$^h$ 19$^m$ 12${\rlap.}^{s}$48;
$\delta$(2000)= $-$20$^\circ$ 47$^\prime$ 27${\rlap.}^{\prime\prime}$42) that is 
very close to the strongest H$_2$O maser. The position of VLA~3 is also located a 
few arcsec from the position of the 1.4~mm emission detected by us 
($\alpha$(2000)= 18$^h$ 19$^m$ 12${\rlap.}^{s}$13; $\delta$(2000)= 
$-$20$^\circ$ 47$^\prime$ 30${\rlap.} ^{\prime\prime}$57; see Fig. ~2) and we 
tentatively propose that they correspond to the same source. Assuming that the radio 
continuum emission from VLA~3 arises from an HII region, it requires a B3 ZAMS star 
to maintain its ionization. All the above data support the idea that the C 
molecular clump harbors a cluster of stars in an early stage of formation.
Mid-IR observations indicate the presence of a pre-main sequence stellar cluster
(Aspin \& Geballe 1992; Yamashita et al. 1987) at this position.

The derived parameters of the C ammonia component, as well as of the other 
ammonia components, are summarized in Table~4. The C ammonia source is the 
warmer component having a rotational temperature of $\sim$ 24 K. It shows a velocity 
gradient of $\sim$3~km~s$^{-1}$ over an angular scale of $\sim$12$^{\prime\prime}$ 
(see Fig. 9). If due to rotation, it implies a mass of $\sim$30~ M$_\odot$, which is 
in good agreement with the mass of $\sim$40 M$_\odot$ estimated from the virial 
assumption (Table~4).
 
\subsection{The NE component}

The emission from the NE ammonia clump is seen in the velocity range
from 10.0 to 10.6 km~s$^{-1}$. Neither HCN nor radio continuum emission was 
detected toward this clump. This component is associated with methanol masers
(coincident within the positional error of the masers of 10$^{\prime\prime}$;
see Figure~3) as well as with weak HCO$^+$ emission. The 95~GHz class I methanol 
masers found are in a velocity range from 12.4 to 13.7 km~s$^{-1}$  (Val'tts et 
al. 2000), slightly redshifted with respect to the LSR velocity of the ammonia clump 
($\sim$11 km~s$^{-1}$). Class I methanol masers are characterized by a lack of radio 
continuum emission, and may be related with the first stages of massive stellar 
evolution (Garay \& Lizano 1999). The relatively warm rotational temperature of this 
component (22~K; see Table 4) also favors the presence of embedded stars.
The molecular mass estimated from the ammonia data is $\sim$ 1 M$_\odot$.
On the other hand, the virial mass is aproximately 13 M$_\odot$, suggesting that 
this molecular clump may not be in equilibrium.
Most probably, this warm clump is a substructure of the C component.
In fact, from the channel maps of the (1,1) ammonia emission (Fig. 5) we 
note that at 11.9 km~s$^{-1}$ there are several weak extended structures 
around the C component. In this channel map we also observe an EW filament 
resembling the "ammonia fingers" in OMC-1 (Wiseman \& Ho 1998).

\subsection{The S component}

This molecular clump was detected only in the (1,1) main transition of NH$_3$ and appears 
located toward the south of the exciting source of the jet. We did not detect emission
of HCO$^+$ and HCO at a 3$\sigma$ level of 1.5 Jy~beam$^{-1}$. 
The ammonia (1,1) line has V$_{LSR}\sim$ 13~km~s$^{-1}$, slightly redshifted with respect
to the C component. The low rotational temperature of this component (10 K; see 
Table 4) suggests that it does not 
have embedded an object of large luminosity. The estimates of the molecular
mass and the virial mass are very similar, $\sim$ 13 M$_\odot$, suggesting that
this clump is probably in equilibrium.
It will then be interesting to study this component in detail since it may
be a precursor to the formation of stars.
Mart\'\i\ et al. (1993) found that along the north-south
direction of the thermal jet there are several knots  that emit at radio frequencies
and in particular the S ammonia component is close to one of those knots (source VLA~13 from
Mart\'\i ~et al. 1993): 21$^{\prime\prime}$ or 0.18 pc in projection. 
No counterpart was found at other frequencies for this new S ammonia clump using the SIMBAD database.
An inspection of the 2MASS All
Sky Survey shows that there are many near-IR sources in the GGD 27
region. Yet, there is a region of roughly 40$^{\prime\prime}$ of diameter around the
S ammonia clump where there are no near-IR sources. This suggests
that the S ammonia clump is part of a very dark molecular cloud, where
apparently there is no star formation going on. The presence of a possible
shock excited radio source nearby (source VLA 13) suggests that this clump may belong to
the class of irradiated clumps, that is, clumps that are exposed to the
strong radiation of a Herbig-Haro object (e.g. Girart et al. 1994, 2002).
In these clumps the ammonia emission is strongly enhanced (Viti \&
Williams 1999; Viti et al. 2003).  If the S clump is an irradiated
clump, the estimated mass would be much lower than 13 M$_\odot$. 

\subsection{Emission Toward the Jet}

Our results show the presence of SO molecular emission and continuum emission
at mm wavelengths in the direction of the exciting source of the HH 80-81 jet.
The 5$_5$--4$_4$ transition of SO is a tracer of warm ($\geq$40~K) and dense 
($\sim$10$^6$~cm$^{-3}$) gas (Chernin et al. 1994).
The SO emission is detected at $>$ 4$\sigma$ level in both the spectrum and the
integrated map of Figure~8. From the SO line emission and assuming optically thin 
conditions, we obtain a beam-averaged column density of N(SO) in the range 2.5 
$-$ 3.4 $\times$ 10$^{14}$ cm$^{-2}$ for a temperature range of 50 to 100 K. 
Assuming that the SO/H$_2$ ratio is in the 0.5 $-$ 2.0 $\times$ 10$^{-8}$ range 
(Helmich \& van Dishoeck 1997), we obtain a beam-averaged
H$_2$ column density in the range of 1.3 $-$ 6.8 $\times$ 10$^{22}$ cm$^{-2}$, that
translates in a mass of 0.7 $-$ 3.5 M$_\odot$.
On the other hand, assuming that the dust emission is optically thin
and that the dust temperature is in the 50 $-$ 100 K range, we derive
a total mass from the dust emission of 3.7 $-$ 1.8 M$_\odot$ (Beckwith \& Sargent 1991).
We conclude that both the SO and the dust data are consistent with
the presence of a few solar masses (1 $-$ 3 M$_\odot$) of gas closely associated
with the surroundings of the exciting source of the jet. 

We found that no strong ammonia emission was detected toward the jet.
The position of the exciting source of the thermal jet lies at the edge of the
C ammonia component (see Figure~4). In the image of the C component 
(Figure~6) we see a small enhancement of the (2,2) ammonia emission along the 
direction of the thermal jet, while there is no emission in the (1,1) image. 
Figure~10 shows the spectra of the (1,1) and (2,2) inversion transitions at the peak
position of the thermal jet. We detect a weak and broad (2,2) ammonia line emission
which suggests that warm ammonia gas is present in the vicinity of the exciting 
source of the thermal jet. From a Gaussian profile fit to the (2,2) spectrum we 
obtain a peak flux of 4.8 $\pm$ 0.7 mJy~beam$^{-1}$, an LSR line center velocity 
of 11.5 $\pm$ 0.3 km~s$^{-1}$ and a line width of 5.4 $\pm$ 0.8
km~s$^{-1}$. Using these parameters for the (2,2) line and a 2$\sigma$ upper limit 
for the (1,1) ammonia line of $\sim$3 mJy~beam$^{-1}$, it is possible to estimate 
a lower limit to the rotational temperature of $\sim$50~K.
This is unusually warm molecular gas, that we suggest is being heated by the
star that produces the jet. However, the determination of its precise location 
and nature cannot be addressed with the data presented here.
A crude upper limit for the molecular mass associated with
the jet can be obtained as follows. From the peak (2,2) line
flux density of 4.8 mJy beam$^{-1}$ and the angular resolution
of the observations, we estimate a line brightness temperature
of $\sim$0.7 K. Assuming that the excitation temperature is 50 K,this implies
$\tau(2,2,m) \simeq 0.014$, an optically thin line. Adopting 
a line width of 5.4 km s$^{-1}$, we derive a total column density 
of ammonia of $N(NH_3) \simeq 2 \times 10^{14}$ cm$^{-2}$.
Using the ratio of [$H_2$/NH$_3$] = 10$^8$ used for the ammonia 
components, this implies $N(H_2) \simeq 2 \times 10^{22}$ cm$^{-2}$.
Finally, assuming that the gas is distributed over a physical size
similar to the beam, we obtain a mass estimate of 0.3 M$_\odot$.
It is unclear if this molecular gas is associated with the
jet or with the edge of the C molecular component.
In any case, the mass derived from ammonia for the surroundings of the
jet is an order of magnitude smaller than that derived from SO or dust, 
suggesting that ammonia is strongly depleted. 

\section{Conclusions}

We have made high angular resolution VLA and BIMA observations of NH$_3$, HCO$^+$, 
HCN, and SO molecular gas as well as 1.4, 3.5 and 7 mm continuum emission toward 
the exciting source of HH~80-81. Our main conclusions are summarized as follows.

We detected millimeter continuum emission (1.4, 3.5 and 7~mm), as well as SO 
5$_5$--4$_4$ line emission, associated with the exciting source of the HH~80-81 
outflow. These emissions may be arising from a circumstellar disk, but higher 
angular resolution data are required to study the morphology of the emitting 
dust and gas. This is the first time that SO 5$_5$--4$_4$ line emission is 
detected toward the exciting source of HH~80-81,  suggesting 
that this transition may be a good tracer of molecular gas near massive protostars.
Only weak ammonia (2,2) emission was detected toward the exciting source of the jet, 
indicating a temperature $\sim$50 K. Unfortunately, the precise location and nature 
of the warm molecular gas traced by the ammonia cannot be determined with the 
present data.

Three molecular components have been detected in the surroundings of the
exciting source of the HH~80-81 outflow. The brightest of these condensations
(the C component) was detected in NH$_3$, HCO$^+$, HCN and appears associated 
with H$_2$O maser emission suggesting that this component harbors a
cluster of young stars. The NE component was detected in NH$_3$ and HCO$^+$ and is 
associated with class I methanol masers. The S component was detected only in 
NH$_3$, and seems to be a massive (about 15 M$_\odot$) and cold clump without 
on-going stellar formation. Alternatively, this component may belong to the class of 
irradiated clumps, with VLA~13 being the source of the radiation.

\medskip

\acknowledgments
YG and LFR acknowledge financial support from DGAPA-UNAM 
and CONACyT, M\'exico.
JMG is supported by MCyT grant AYA2002-00205.
GG acknowledges support from FONDECYT project 1010531.
JM also acknowledges partial support by DGI of the
Ministerio de Ciencia y Tecnolog\'{\i}a (Spain) under grant AYA2001-3092,
as well as by Junta de Andaluc\'{\i}a (Spain) under PAI group FQM322.
This research has made use of the SIMBAD database,
operated at CDS, Strasbourg, France.
The 2MASS project is a collaboration between The University of Massachusetts
and the infrared processing Analysis Center (JPL/Caltech), with funding 
provided primarily by NASA and NSF.

\clearpage
%
\begin{deluxetable}{lrccrccc}
\tablewidth{0pt}
\tablecaption{PARAMETERS OF THE BIMA OBSERVATIONS 
\label{tbima}}
\tablehead{
&
&
\colhead{$\nu$}&
\multicolumn{2}{c}{Synthesized Beam}&
\colhead{Bandwidth}&
\colhead{$\Delta v$\tablenotemark{a}}&
\colhead{$RMS$\tablenotemark{b}}
\\
\cline{4-5}
\colhead{Epoch}&
\colhead{Observation}&
\colhead{(GHz)}&
\colhead{HPBW}&
\colhead{P.A.}&
\colhead{(MHz)}&
\colhead{(km\,s$^{-1}$)}&
\colhead{(Jy\,Beam$^{-1}$)}
}
\startdata
1999 Sep 27 
&Continuum  
 &85.12    &$19\farcs0\times6\farcs3$&$-14\degr$ &100 &    &0.021 \\
& HCO$^+$ 1--0 
 &89.18852 &$18\farcs7\times6\farcs1$&$-14\degr$ &25  &0.66&0.56  \\
& HCN 1--0   
 &88.63185 &$18\farcs3\times6\farcs1$&$-14\degr$ &25  &0.66&0.52  \\
2001 May 6 
&Continuum\tablenotemark{c}  
 &216.77   & $7\farcs4\times2\farcs9$& $-7\degr$ &1200&    &0.038 \\
& SO 5$_5$--4$_4$   
 &215.22065& $7\farcs4\times3\farcs1$& $-9\degr$ &25  &0.68&1.80  \\
& H$_2$CO 3$_{0,3}$--2$_{0,2}$   
 &218.22219& $7\farcs4\times3\farcs1$& $-9\degr$ &25  &0.67&2.00  \\
\enddata
\tablenotetext{a}{Spectral resolution.}
\tablenotetext{b}{Per channel. The brightness temperature of the 3.4~mm data
is $T_b(K) \simeq 1.4 \times 10^{-3} S(mJy)$, while for the 1.4~mm data is
$T_b(K) \simeq 1.1 \times 10^{-3} S(mJy)$.}
\tablenotetext{c}{The continuum maps at 1.4~mm were made using the whole upper and
lower sidebands.}
\end{deluxetable}

\clearpage

\begin{deluxetable}{lrr}
\tablewidth{0pt}
\tablecaption{DUST EMISSION
\label{tdust}}
\tablehead{
\colhead{$\lambda$}&
\multicolumn{2}{c}{Total Flux (Jy)}
\\
\cline{2-3}
\colhead{(mm)}&
\colhead{Thermal Jet}&
\colhead{C component}
}
\startdata
7.0 & 0.014$\pm$0.002 &    $\la 0.006$    \\
3.5 & 0.10$\pm$0.02   &    $\la 0.08$ \\
1.4 & 0.97$\pm$0.11   & 0.21$\pm$0.05 \\
\enddata
\end{deluxetable}

\clearpage

\begin{deluxetable}{lccccccc}
\footnotesize
\tablecaption{AMMONIA (1,1) AND (2,2) LINE PARAMETERS \label{tbl-2}}
\tablehead{
&   Peak Position & & \\
\cline{2-3}
Component & $\alpha$(2000)& $\delta$(2000) & V$_{LSR}$ & $\Delta$V &{Flux Density$^a$} & 
{$\Theta_S ^b$} \\
&  (h~~m~~s) &($\circ ~~\prime ~~{\prime\prime}$) & (km~s$^{-1}$) & (km~s$^{-1}$) & (mJy) & 
(${\prime\prime}$) \\
}
\startdata
C~~~~~ (1,1;m) & 18 19 12.54  & -20 47 25.5 & 12.2 $\pm$ 0.1 & 2.3 $\pm$ 0.1 & 260 $\pm$ 6 & 8.9 \\
~~~~~~~ (1,1;s) &              &             & 12.1 $\pm$ 0.1 & 2.6 $\pm$ 0.1 &  80 $\pm$ 6 & \\
~~~~~~~ (1,1;s) &              &             & 12.2 $\pm$ 0.1 & 1.9 $\pm$ 0.1 & 108 $\pm$ 6 &  \\
C~~~~~ (2,2;m) & 18 19 12.55  & -20 47 25.4 & 12.2 $\pm$ 0.1 & 2.4 $\pm$ 0.1 & 188 $\pm$ 6 & 10.6 \\
NE~~~ (1,1;m) & 18 19 13.52  & -20 47 13.2 & 10.6 $\pm$ 0.1 & 2.1 $\pm$ 0.2 & 25  $\pm$ 2 & 3.5 \\
~~~~~~~ (1,1;s) &              &             & 11.7 $\pm$ 0.4 & 3.4 $\pm$0.8  & 8 $\pm$ 2 & \\
~~~~~~~ (1,1;s) &              &             & 11.3 $\pm$ 0.4 & 3.9 $\pm$0.9  & 8 $\pm$ 2 & \\
NE~~~ (2,2;m) & 18 19 13.42      & -20 47 13.1   & 10.5 $\pm$ 0.1 & 1.9 $\pm$ 0.3 & 14  $\pm$ 2 & \\
S~~~~~ (1,1;m) & 18 19 11.94  & -20 48 29.3 & 13.0 $\pm$ 0.1 & 1.6 $\pm$ 0.1 & 111 $\pm$ 6 & 6.1 \\
~~~~~~~ (1,1;s) &              &             & 12.9 $\pm$ 0.1 & 1.6 $\pm$ 0.2 & 67 $\pm$ 6 & \\
~~~~~~~ (1,1;s) &              &             & 13.0 $\pm$ 0.1 & 1.7 $\pm$ 0.2 & 58 $\pm$ 6 & \\
S~~~~~ (2,2;m) &   &  &  & & $<$ 18 &  \\
\enddata
\tablenotetext{a}{At the peak of the line.} 
\tablenotetext{b}{Angular size estimated from a Gaussian fit.}
\end{deluxetable}

\clearpage
\begin{deluxetable}{lcccccccc}
\footnotesize
\tablecaption{DERIVED PARAMETERS OF THE AMMONIA STRUCTURES \label{tbl-4}}
\tablehead{
Component& $\tau_m$(1,1)& $\tau_m$(2,2) & {\it l~$^a$} & T$_{rot}$ & N(NH$_3$) & n(H$_2$)$^b$ & M(H$_2$)& M$_{vir}$ \\
& & & (pc) & (K) & (cm$^{-2}$) & (cm$^{-3}$) & (M$_\odot$) & (M$_\odot$)\\
}
\startdata
C& 0.77 & 0.49 & 0.073 & 24 & 2.5 $\times$ 10$^{15}$ & 1 $\times$ 10$^6$ & 16& 40 \\
NE &     0.39 & 0.20 & 0.029 & 21 & 1.0 $\times$ 10$^{15}$ & 1 $\times$ 10$^6$ & 1 & 13 \\
S  & 2.63 & $<$0.16 & 0.050 & $<$10 & 4.9 $\times$ 10$^{15}$ & 3 $\times$ 10$^6$ & 15 & 13 \\
\enddata
\tablenotetext{a}{Diameter estimated assuming a distance of 1.7 kpc.}
\tablenotetext{b}{Derived assuming $[H_2/NH_3] = 10^8$}
\end{deluxetable}

\clearpage

\begin{figure*}
\includegraphics[width=1.0\textwidth,angle=0]{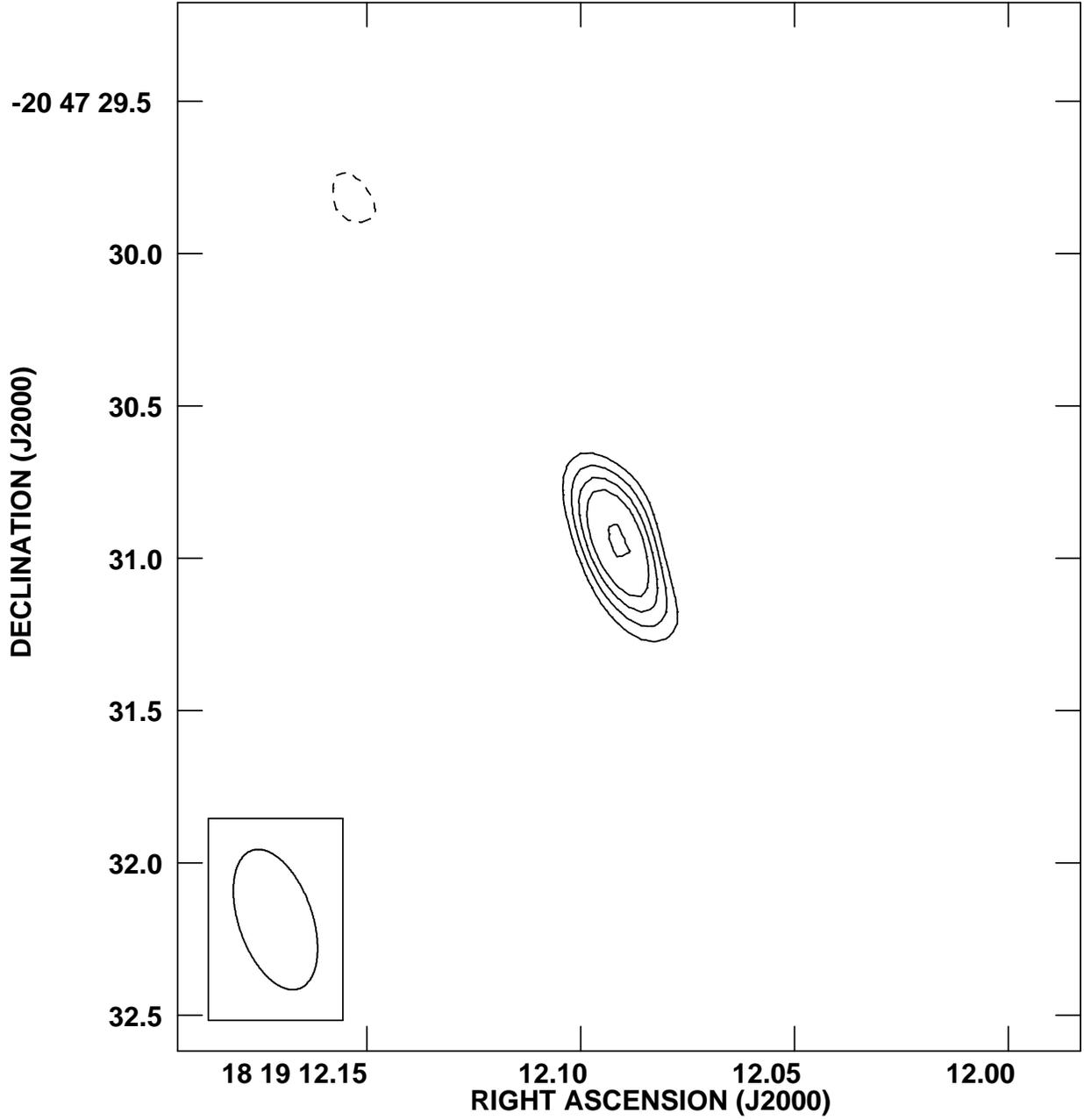}
\caption[f1.ps]{VLA image made at 7 mm of the exciting source of the HH~80-81
outflow.  Contours are $-$3, 3, 4, 5, 6, and 8 $\times$ 1.3 mJy beam$^{-1}$,
the $rms$ noise of the image.
\label{fig-1}}
\end{figure*}

\clearpage

\begin{figure}[hbt]
\epsscale{0.70}
\plotone{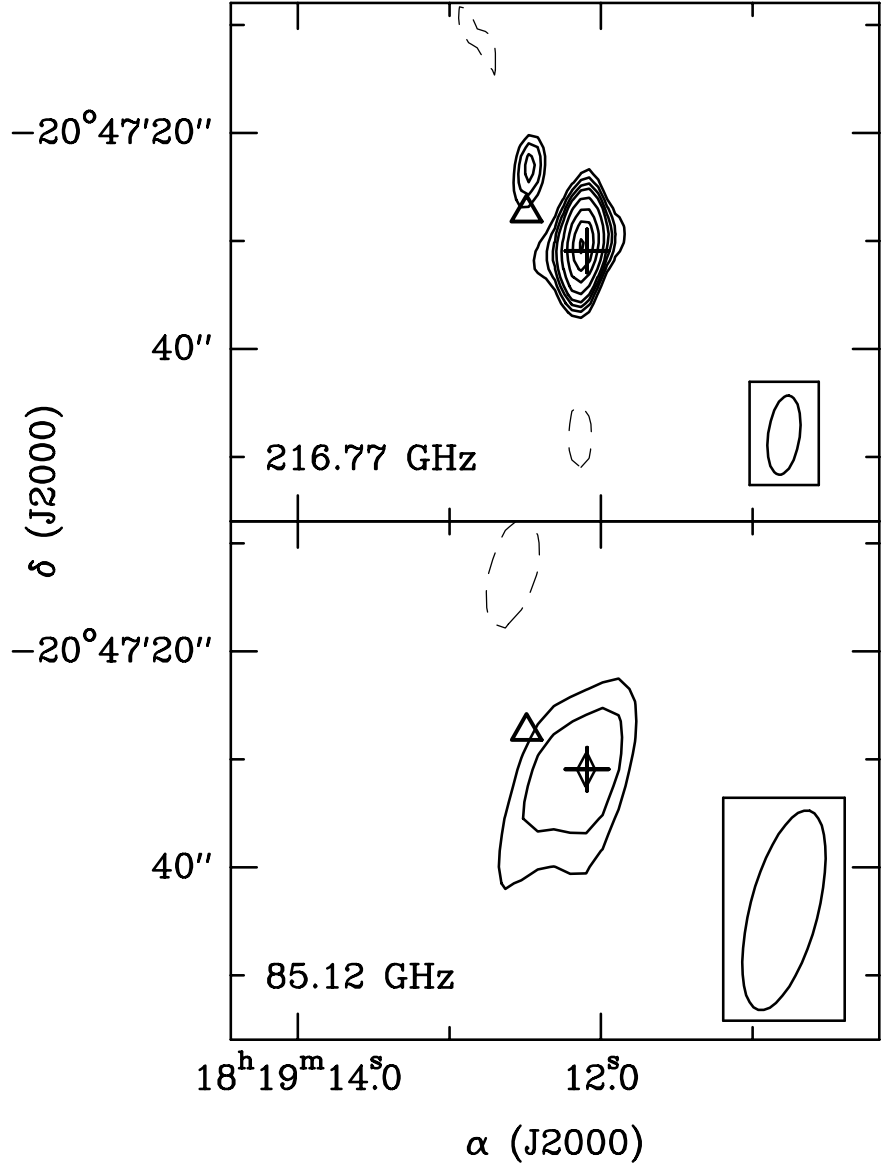}
\figcaption{Images of the BIMA continuum emission at 1.4~mm (top panel) and
3.5~mm (bottom panel). The contour levels are $-$3, 3, 4, 5, 6 8, 11, 14 and 17
$\times$ 0.038 and 0.021~Jy~beam$^{-1}$, the {\em rms} noise of the 1.4~mm and
3.5~mm maps, respectively. The cross indicates the position of the thermal jet
and the triangle marks the position of the H$_2$O maser. 
The synthesized beams are shown in the bottom right corner of each panel.
\label{fig-2}
}
\end{figure}

\clearpage

\begin{figure*}
\includegraphics[width=1.0\textwidth,angle=0]{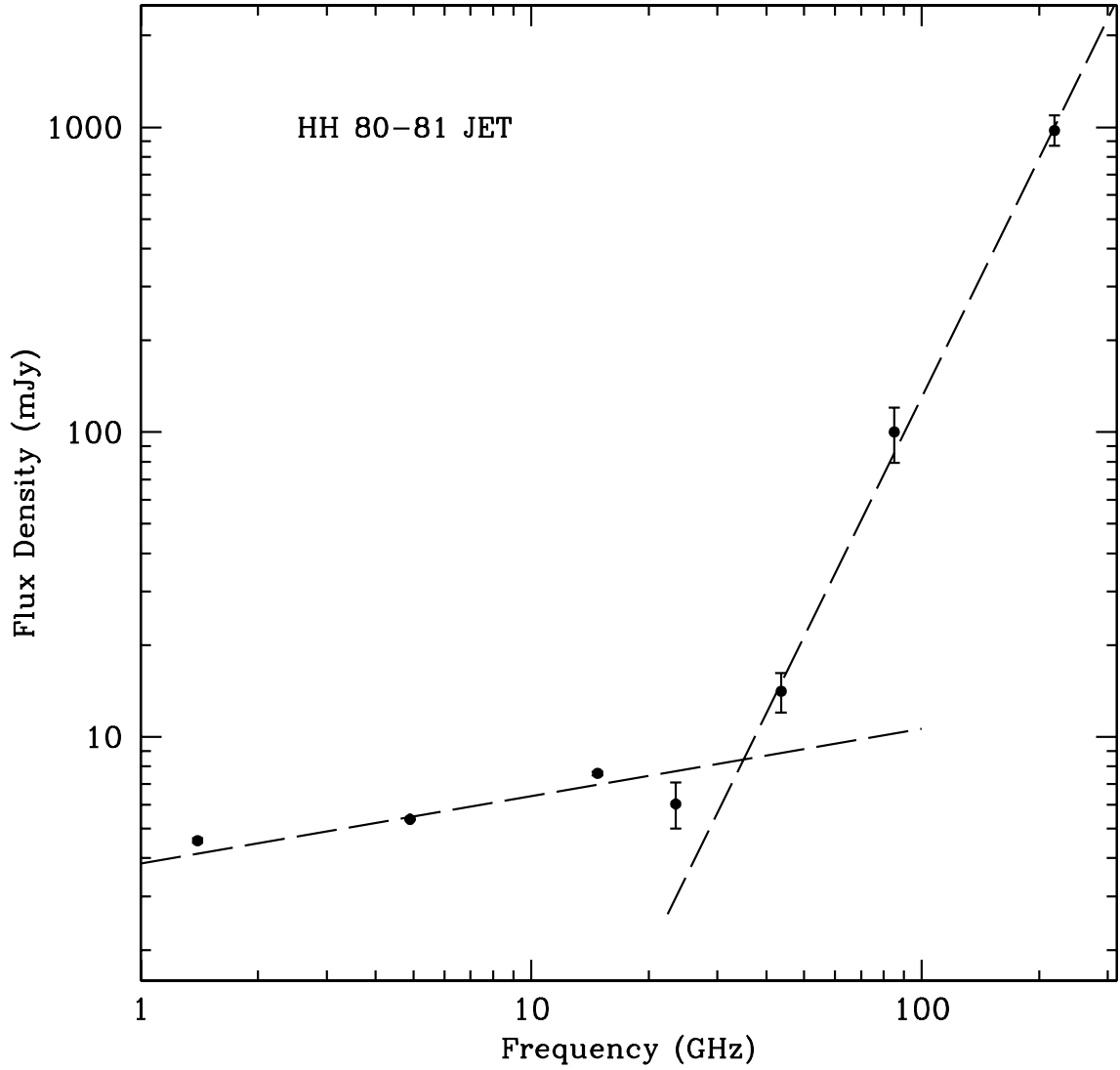}
\caption[f3.eps]{Continuum cm and mm spectrum of the region
associated with the exciting source.
The dashed line to the left is a linear fit to the four lower frequency data points,
while the dashed line to the right is a linear fit to the three higher frequencies
data points.
\label{fig-3}}
\end{figure*}

\clearpage

\begin{figure*}
\includegraphics[width=1.0\textwidth,angle=0]{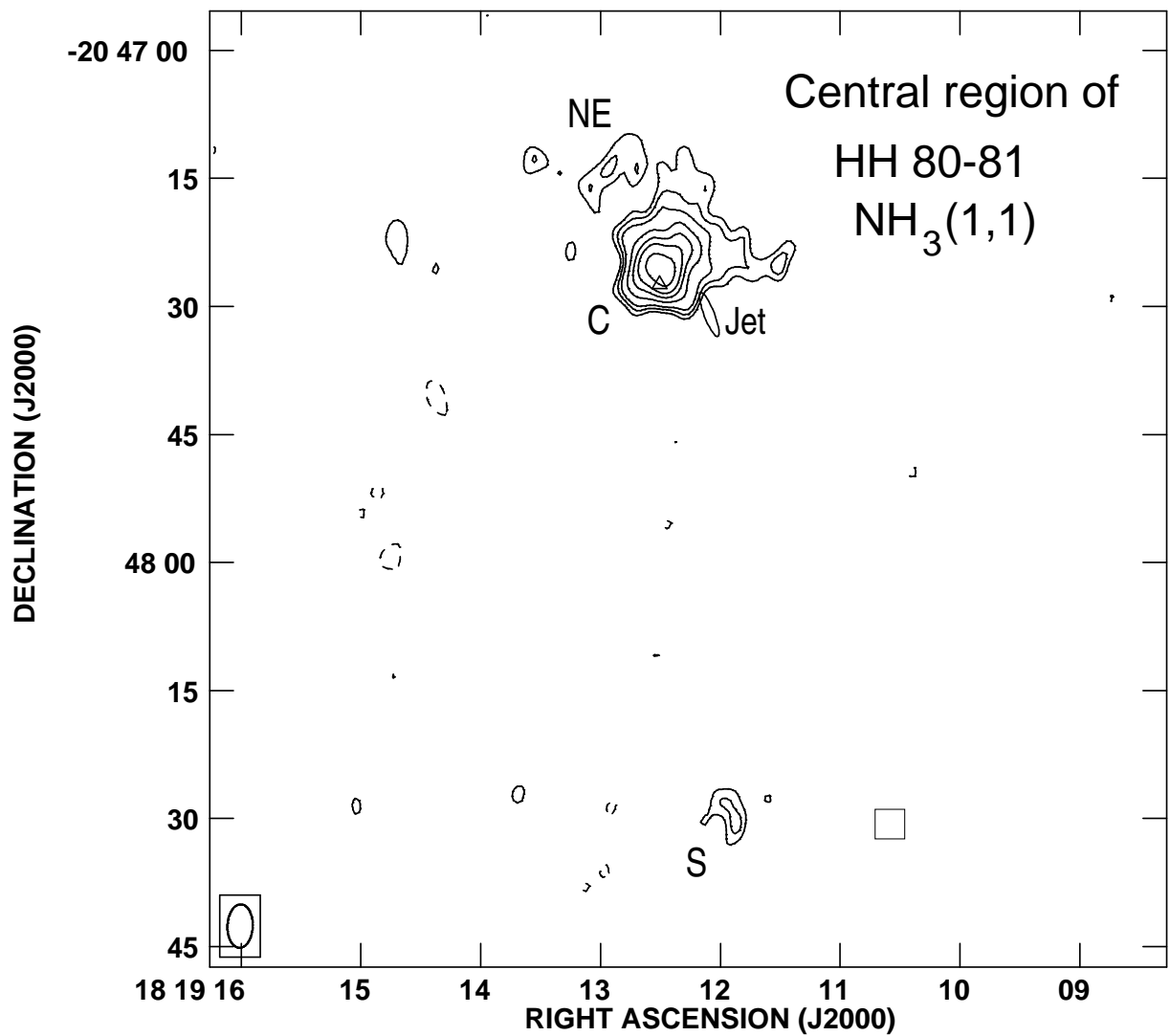}
\caption[f4.ps]{Image of the NH$_3$ (1,1) 
velocity-integrated 
(from 10.0 to 14.3 km s$^{-1}$)
line emission from the surroundings of the
exciting source of the HH 80-81 region. Contour levels 
are -3, 3, 4, 5, 7, 9, 11 and 14 $\times$ 1.6 
mJy beam$^{-1}$, the $rms$ noise of the image. This image 
is not corrected for the response of
the primary beam and the S component appears weaker than 
it really is. The parameters given
in Table~3 are corrected for the primary beam response. The ellipse indicates 
the position and
alignment of the thermal jet, the triangle marks the H$_2$O maser position 
taken from G\'omez et al. (1995) and the small square to the south 
indicates the position of the radio continuum source 
VLA~13 corresponding to a knot of the jet (Mart\'\i ~et al. 1993).
\label{fig-4}}
\end{figure*}

\clearpage 

\begin{figure*}
\includegraphics[width=0.80\textwidth,angle=0]{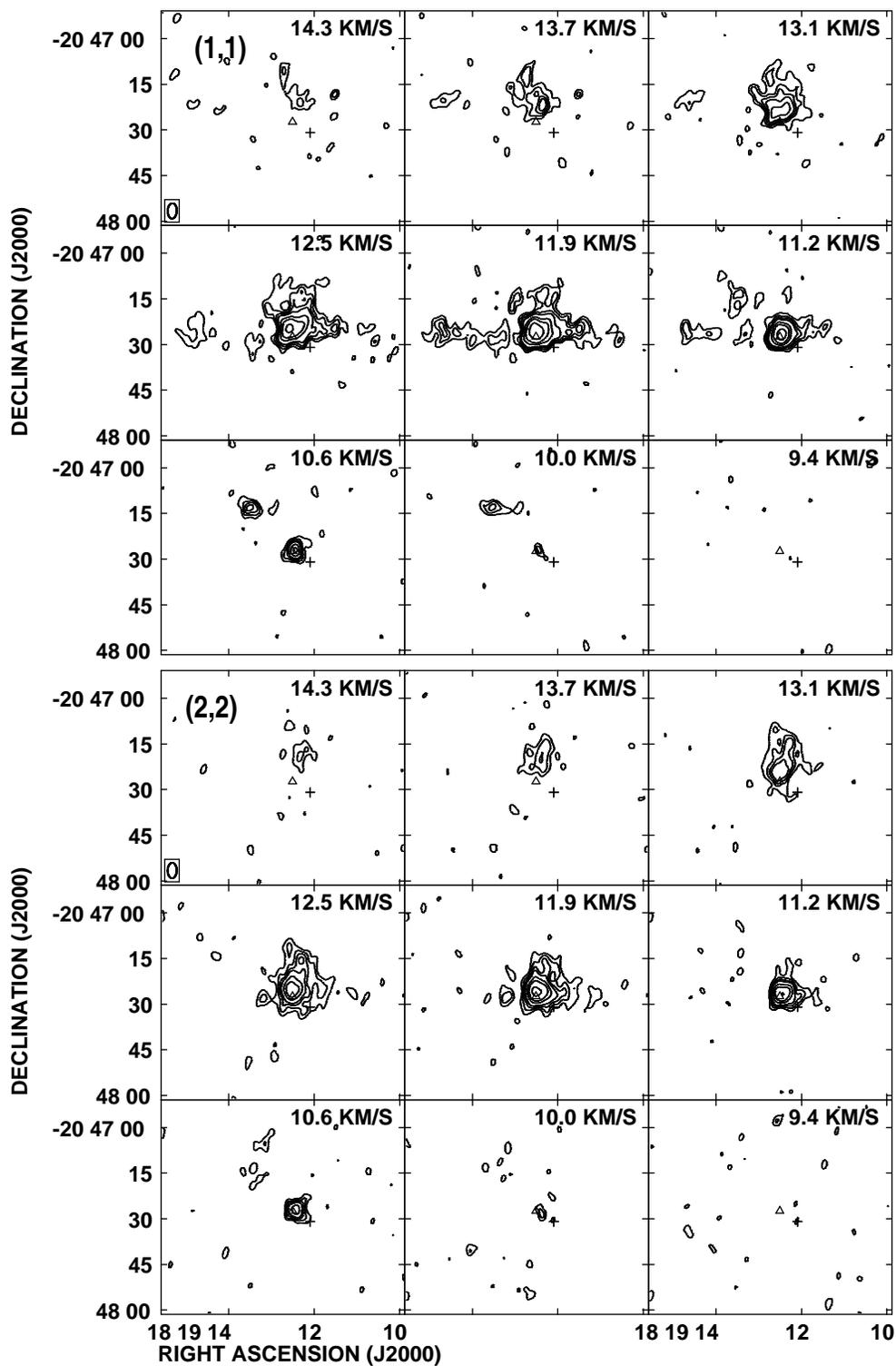}
\caption[f5.ps]{ Channel maps of the (1,1) (top) and 
(2,2) (bottom) inversion transition lines 
of ammonia toward the HH~80-81 region. Contour levels 
are 4, 6, 8, 10, 15, 20, 30 and 50 $\times$
1.5 and 1.4 mJy beam$^{-1}$, respectively.
The angular resolution is 5${\rlap.}^{\prime\prime}$1 $\times$ 3${\rlap.}^{\prime\prime}$0 at 
P.A.= $-$2$^\circ$. The small cross indicates the position of the thermal jet and the
triangle marks the position of the H$_2$O maser. 
\label{fig-5}}
\end{figure*}

\clearpage

\begin{figure}
\includegraphics[width=0.35\textwidth,angle=0]{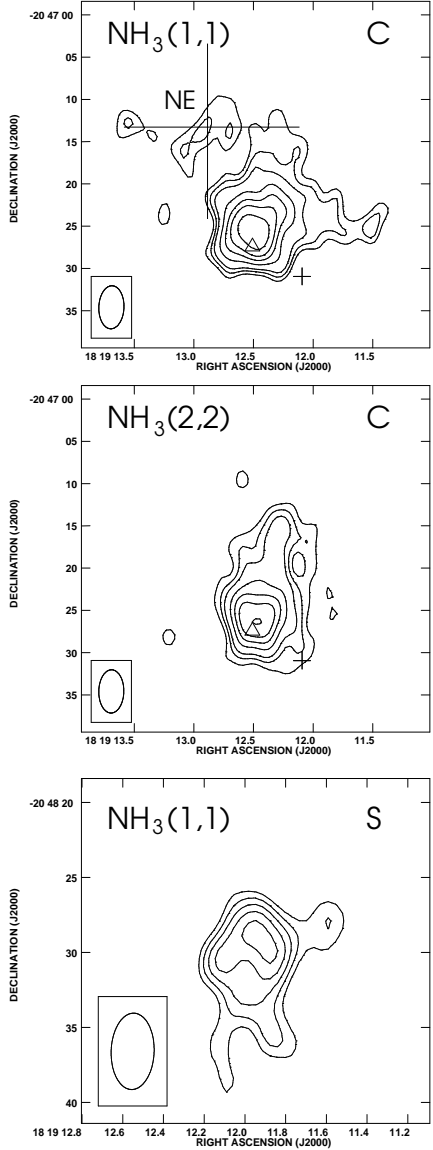}
\caption[f6.ps]{Velocity integrated images of the new
ammonia components; C, NE and S components. (Top and middle) C and NE clumps
in the (1,1) and (2,2) ammonia emission integrated in a velocity range from 
10.0 to 14.3 km~s$^{-1}$. The contour levels are -3, 3, 4, 5, 7, 9, 11 and 14
$\times$ 1.6 and 1.5 mJy beam$^{-1}$, the $rms$ noise of the (1,1) and (2,2) 
maps respectively. The small cross indicates the position of the thermal jet, the
triangle marks the position of the H$_2$O masers and the large cross indicates 
the position and error of the methanol masers taken from Val'tts et al. (2000).
(Bottom) Ammonia  (1,1) emission of the S component integrated
in a velocity range from 11.9 to 13.7 km~s$^{-1}$. The contour levels are -4, 4, 5,
6, 7, and 8 $\times$ 3.0 mJy beam$^{-1}$, the $rms$ noise of the map. 
This image was corrected for the response of the primary beam. The S component was
detected only in the (1,1) ammonia transition. 
\label{fig-6}}
\end{figure}

\clearpage

\begin{figure*}
\includegraphics[width=1.0\textwidth,angle=0]{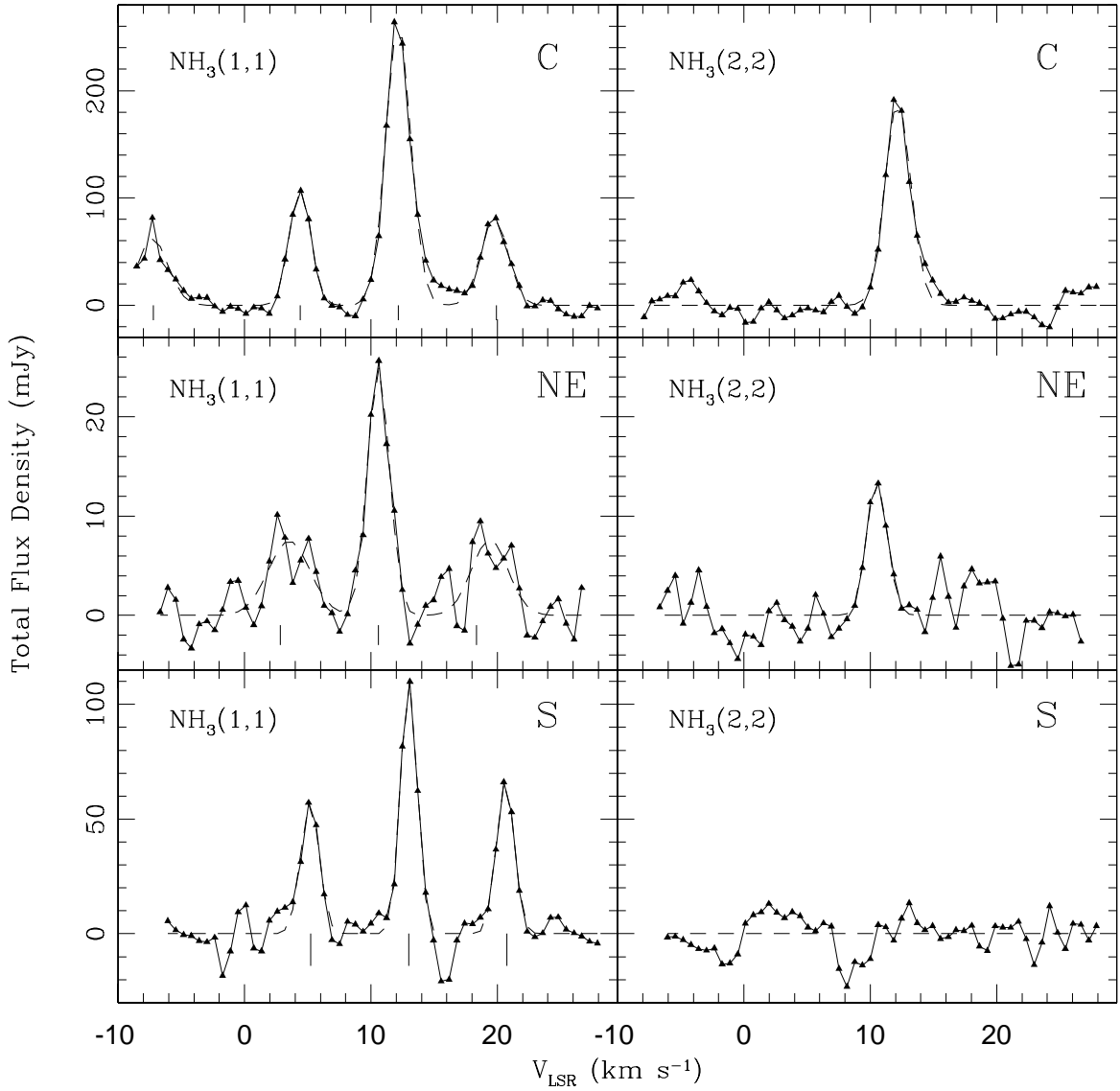}
\caption[f7.ps]{Integrated spectra in the (1,1) and (2,2) lines of NH$_3$
from the different molecular components in the HH~80-81 region. The small vertical bars indicate the
expected velocities of the main and satellite (1,1) ammonia lines. 
\label{fig-7}}
\end{figure*}

\clearpage

\begin{figure}[hbt]
\plotone{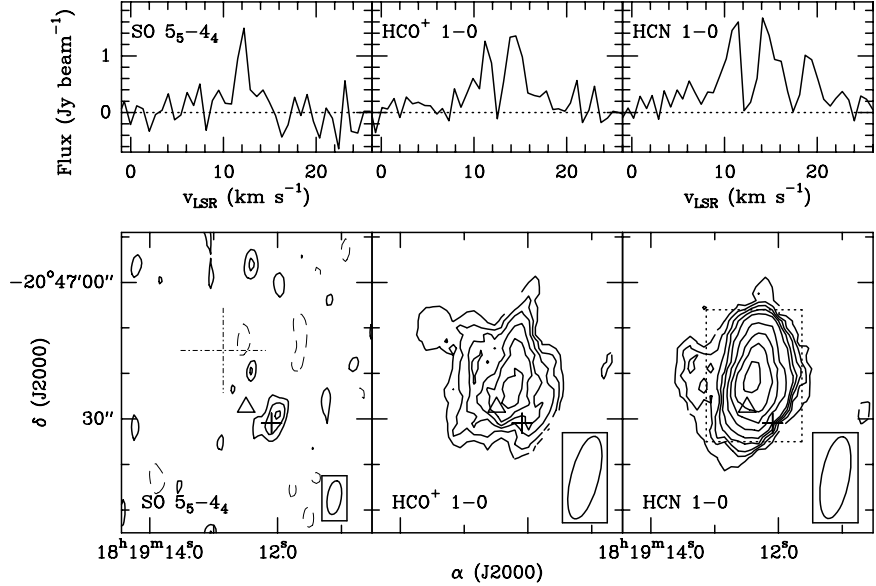}
\figcaption{
{\em Bottom panels}. Images of the BIMA integrated emission of the SO
5$_5$--4$_4$ (left panel), HCO$^+$ 1--0 (central panel), HCN 1--0 (right panel)
lines. The SO map shows the integrated emission in a velocity range from 10.0
to 14.3 km~s$^{-1}$. The HCO$^+$ and HCN maps show the integrated emission in a
velocity range from 9.5 to 15.7 km~s$^{-1}$.  The contour levels are $-$2, 2,
3, 4, 5, 6, 8, 10, 12 and 14 $\times$ 3.2, 0.15,
0.13~Jy~beam$^{-1}$, the {\em rms} noise of the SO, HCO$^+$ and HCN
maps, respectively.  The small cross indicates the position of the exciting source
of the thermal jet and the triangle marks the position of the H$_2$O maser. 
The large cross indicates the position and error of the methanol masers taken
from Val'tts et al. (2000).
The synthesized beams are shown in the bottom right corner of each panel. 
{\em Top panels}. Spectra of  the SO 5$_5$--4$_4$ (left panel), HCO$^+$ 1--0
(central panel), HCN 1--0 (right panel) lines. The SO spectrum was taken at the
position of the thermal radio jet and divided by a factor of 5 in order to
put it at the same scale than the HCO$^+$ and HCN spectra that are weaker than the SO. 
For the HCO$^+$ and
HCN the spectra was obtained from the averaged emission within the dashed box
shown in the HCN map.
\label{fig-8}
}   
\end{figure}

\clearpage

\begin{figure*}
\includegraphics[width=0.35\textwidth,angle=0]{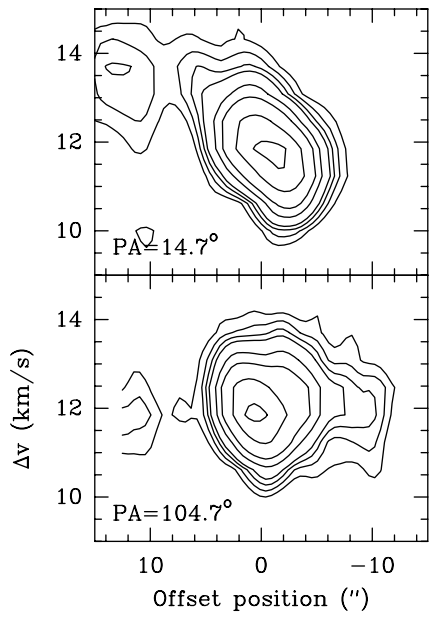}
\caption[f9.ps]{Position-velocity diagrams of the NH$_3$ (1,1) emission
from the C component along  the major axis direction at P.A.=14.7$^\circ$ (top) 
and along the minor axis direction P.A.=104.7$^\circ$ passing through $\alpha$(2000)= 18$^h$ 19$^m$
12${\rlap.}^s$5; $\delta$(2000)= $-$20$^\circ$ 47$^\prime$ 24${\rlap.}^{\prime\prime}$9. 
The contour levels are $-$4, 4, 6, 8, 10, 15, 20, 30, and 40 $\times$ 1.4 mJy~beam$^{-1}$.
\label{fig-9}}
\end{figure*}

\clearpage

\begin{figure*}
\includegraphics[width=1.0\textwidth,angle=0]{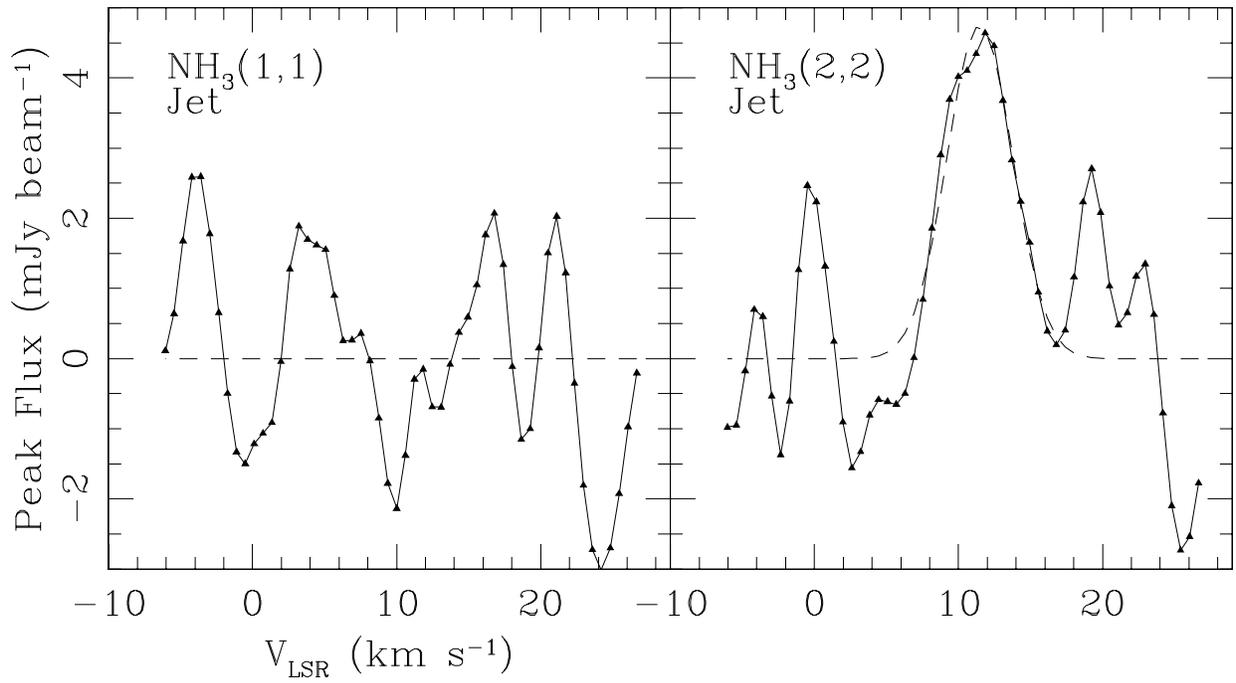}
\caption[f10.ps]{Ammonia spectra in the (1,1) and (2,2) lines 
at the peak position of the HH~80-81 thermal jet: $\alpha$(2000)= 18$^h$ 19$^m$ 
12${\rlap.}^s$509; $\delta$(2000)= $-$20$^\circ$ 47$^\prime$ 
27${\rlap.}^{\prime\prime}$42. Only weak and broad (2,2) emission was detected.   
\label{fig-10}}
\end{figure*}


\clearpage

\begin{thebibliography}{}

\bibitem[Aspin \& Geballe 1992]{AG92} Aspin, C., \& Geballe, T. R. 1992,
\aap, 266, 219

\bibitem[Beckwith \& Sargent 1991]{BS91} Beckwith, S. V. W., \& Sargent,
A.~I. 1991, ApJ, 381, 250 

\bibitem[Beckwith \& Sargent 1996]{BS96} Beckwith, S. V. W., \& Sargent, 
A.~I. 1996, Nature, 383, 139

\bibitem[Bonnell \& Bate 2002]{BB02} Bonnell, I. A., \& Bate, M. 2002,
MNRAS, 336, 659

\bibitem[Cesaroni et al. 1997]{C97} Cesaroni, R., Felli, M., Testi, L.,
Walmsley, C. M., \& Olmi, L. 1997, \aap, 325, 725

\bibitem[Chernin et al. 1994]{CH94} Chernin, L. M., Masson, C. R., \&
Fuller, G. A. 1994, \apj, 436, 741

\bibitem[Devine et al. 1999]{Dev99} Devine, J. D., Bally, J., Reipurth, B.,
Shepherd, D., \& Watson, A. 1999, \aj, 117, 2919

\bibitem[Garay \& Lizano 1999]{GL99} Garay, G., \& Lizano, S. 1999, PASP, 
111, 1049

\bibitem[Garay, et al. 2003]{Gar03} Garay, G., Brooks, K., Mardones, D.,
\& Norris, R. P.  2003, \apj, 587, 739 

\bibitem[Girart et al. 1994]{GRA94} Girart, J. M., et al. 1994, \apj, 435, L145.

\bibitem[Girart et al. 2001]{GIR01} Girart, J. M., Estalella, R., Viti, S., Williams, D. A.,
\& Ho, P. T. P. 2001, ApJ, 562, L91

\bibitem[Girart et al. 2002]{GIR02} Girart, J. M., Viti, S., Williams, D. A.,
Estalella, R., \& Ho, P. T. P. 2002, A\&A, 388, 1004
  
\bibitem[G\'omez et al. 1995]{GRM95} G\'omez, Y., Rodr\'\i guez, L.~F., \&
Mart\'\i , J. 1995, \apj, 453, 268 

\bibitem[Gyulbudaghian et al. 1978]{GGD78} Gyulbudaghian, A. L., Glushkov, Y. L., \& Denisyuk,
E. K. 1978, ApJ, 224, L137

\bibitem[Heathcote, Reipurth \& Raga 1998]{HRR98} Heathcote, S.,  Reipurth, B.,
 \& Raga, A.~C.  1998, AJ, 116, 1940

\bibitem[Helmich \& van Dishoeck 1997]{HD97} Helmich, F. P., \& van Dishoeck, E. F. 1997,
A\&AS, 124, 205

\bibitem[Kurtz 2000]{KUR00} Kurtz, S. 2000, RMxAAC, 9, 169

\bibitem[Mart\'\i ~ et al. 1993]{MRR93} Mart\'\i , J., Rodr\'\i guez, L. F., 
\& Reipurth, B. 1993, \apj, 416, 208

\bibitem[Mart\'\i ~ et al. 1995]{MRR95} Mart\'\i , J., Rodr\'\i guez, L. F.,
\& Reipurth, B. 1995, \apj, 449, 184 

\bibitem[Mart\'\i ~ et al. 1998]{MRR98} Mart\'\i , J., Rodr\'\i guez, L. F., 
\& Reipurth, B. 1998, \apj, 502, 337 

\bibitem[Mart\'\i ~ et al. 1999]{MRR99} Mart\'\i , J., Rodr\'\i guez, L. F.,
\& Torrelles, J. M. 1999, \aap, 345, L5

\bibitem[Reipurth \& Graham 1988]{RG88} Reipurth, B., \& Graham, J. A. 1988,
\aap, 202, 219 

\bibitem[Rodr\'\i guez et al. 1978]{RMD78} Rodr\'\i guez, L.~F., Moran, J.~M.,
Dickinson, D.~F., \& Gyulbudaghian, A.~L. 1978, \apj, 226, 115

\bibitem[Rodr\'\i guez et al. 1980]{RMH80} Rodr\'\i guez, L.~F., Moran, J.~M.,
Ho, P.~T.~P., \& Gottlieb, W. 1980, \apj, 235, 845

\bibitem[Rodr\'\i guez et al. 1993]{R93} Rodr\'\i guez, L.~F., Mart\'\i, J.,
Cant\'o, J., Moran, J. M., \& Curiel, S. 1993, RexMexA\&A, 25, 23

\bibitem[Rodr\'\i guez et al. 1994]{R94} Rodr\'\i guez, L. F., Garay, G.,
Curiel, S., Ram\'\i rez, S., Torrelles, J. M., G\'omez, Y., \& Vel\'azquez, A. 1994,
\apj, 430, L65

\bibitem[Rodr\'\i guez et al. 1998]{R98} Rodr\'\i guez, L.~F., et al. 1998,
Nature, 395, 355

\bibitem[Rodr\'\i guez et al. 1999]{RAC99} Rodr\'\i guez, L. F.,
Anglada, G., \& Curiel, S. 1999, ApJS, 125 427

\bibitem[Scoville \& Kwan 1976]{SK76} Scoville, N. Z., \& Kwan, J. 1976, ApJ, 206, 718

\bibitem[Shepherd et al. 1998]{S98} Shepherd, D. S., Watson, A. M., Sargent, A. I., 
\& Churchwell, E. 1998, \apj, 507, 861

\bibitem[Shepherd, Claussen \& Kurtz  2001]{SCK01} Shepherd, D. S., Claussen, M. J.,
 \& Kurtz, S. E. 2001, Science, 292, 151 

\bibitem[Torrelles, et al. 1986]{TOR86} Torrelles, J. M., Ho, P. T. P., Moran, J. M., 
Rodr\'\i guez, L. F., \& Cant\'o, J. 1986, ApJ, 307, 787

\bibitem[Torrelles, et al. 1996]{TOR96} Torrelles, J. M., G\'omez, J. F., Rodr\'\i guez, L. F.,
Curiel, S., Ho, P. T. P., \& Garay, G. 1996, ApJ, 457, L107

\bibitem[Trinidad, et al. 2003]{TRI03} Trinidad, M. A., et al. 2003, ApJ, 589, 386

\bibitem[Val'tts et al. 2000]{VES00} Val'tss, I. E., Ellingsen, S. P., Slysh, V. I., 
Kalenskii, S. V., Otrupcek, R., \&  Larionov, G. M. 2000, MNRAS, 317, 315

\bibitem[Viti \& Williams 1999]{VW99} Viti, S., \& Williams, D. A. 1999, MNRAS, 310, 517

\bibitem[Viti et al. 2003]{Vit03} Viti, S., Girart, J. M., Garrod, R., Williams, D. A., \&
Estalella, R. 2003, A\&A, 399, 187

\bibitem[Wilner, et al. 1996]{WHR96} Wilner, D. J., Ho, P. T. P., \& Rodr\'\i guez, L. F. 1996, \apj
470, L117
 
\bibitem[Wilner, et al. 1999]{WRM99} Wilner, D.~J., Reid, M.~J., \& Menten, K.~M. 1999, \apj, 513, 775

\bibitem[Wilner, et al. 2000]{WHKR00} Wilner, D. J., Ho, P. T. P., Kastner, J. H., \&
Rodr\'\i guez, L. F. 2000, \apj, 534, L101

\bibitem[Wiseman \& Ho 1998]{WH98} Wiseman, J. J., \& Ho, P.T.P. 1998, ApJ, 502, 676

\bibitem[Yamashita et al. 1989]{Yam89} Yamashita, T., Suzuki, H., Kaifu, N., Tamura, M.,
Mountain, C. M., \& Moore, T. J. T. 1989, \apj, 347, 894

\bibitem[Zhang, Hunter \& Sridharan 1998]{ZHS98} Zhang, Q., Hunter, T. R., \& 
Sridharan, T. K. 1998, \apj, 505, L151

\bibitem[Zhang, et al. 2002]{ZHSH02} Zhang, Q., Hunter, T. R., Sridharan, T. K., \&
Ho, P. T. P. 2002, \apj, 566, 982

\end{thebibliography}
\end{document}